\documentclass[prl,twocolumn,epsf,psfig]{revtex4}
\usepackage{graphicx}
\DeclareGraphicsExtensions{.pdf,.png,.gif,.jpg}
\usepackage{float}
\usepackage{subfig}
\usepackage{epsfig}
\usepackage{wrapfig}
\usepackage{bbold}
\usepackage{amsmath}

\restylefloat{figure}

\begin{document}

\title{Superconducting and correlated phases of an effective Hubbard model on the BCC lattice}

\author{Theja N. De Silva}
\affiliation{Department of Physics and Biophysics,
Augusta University, Augusta, Georgia 30912, USA.}

\begin{abstract}
We investigate the electronic phases of an effective Hubbard model on the body-centered-cubic lattice, motivated by alkali-doped fulleride molecular solids. The model incorporates renormalized on-site interactions and an effective inverted Hund’s coupling originating from electron–phonon interactions. To access complementary interaction regimes, we employ two theoretical approaches. In the intermediate-coupling regime, the on-site repulsive interaction is approximated by a long-range interaction in momentum space, yielding an exactly solvable Hatsugai–Kohmoto model supplemented by a BCS-type pairing term. Within this framework, we analyze the superconducting instability and demonstrate a first-order normal–superconducting phase transition, characterized by a discontinuous jump of the order parameter. In the strong-coupling regime, where pairing fluctuations are suppressed, we apply the spin-rotationally invariant slave-boson formalism to map out the temperature–interaction phase diagram. This analysis reveals first-order transitions between a Fermi-liquid phase, an antiferromagnetic phase, and a Mott insulating phase, with a narrow intermediate region where all three phases compete. The resulting phase diagram captures the interplay of itinerancy, magnetic order, and Mott localization in three dimensions and provides a unified perspective on superconducting and correlation-driven phenomena in fulleride-inspired lattice systems.
\end{abstract}

\maketitle

\section{I. Introduction}

Alkali-doped fullerides A$_3$C$_{60}$, doped with A = K, Rb, Cs represent a unique class of molecular solids in which strong electron correlations, multiorbital physics, and electron–phonon interactions coexist on comparable energy scales. Since the discovery of superconductivity in K$_3$C$_{60}$, with a transition temperature approaching 40 K, these materials have attracted tremendous interest as the highest critical temperature molecular superconductors known to date~\cite{ref1}. Their narrow conduction bands, derived from the triply degenerate molecular orbitals of C$_{60}$, make fullerides particularly susceptible to correlation-driven phenomena, including Mott localization and magnetic ordering~\cite{ref2, ref2a, ref2b}.

A remarkable feature of alkali-doped fullerides is the existence of distinct crystalline structures, most prominently the face-centered-cubic (FCC) and body-centered-cubic (BCC or A15-type) lattices. In these structures, the C$_{60}$ molecules form a weakly bonded lattice with alkali ions occupying interstitial sites, leading to tunable interactions by changing the lattice spacing by chemical substitution through internal pressure. Experimental studies shows that both FCC and BCC fullerides exhibit rich phase diagrams in which superconductivity emerges in close proximity to Mott insulating and antiferromagnetic phases. In particular, expanded lattices such as Cs$_3$C$_{60}$,  realize a Mott insulating state at ambient pressure, which undergoes a first-order transition to a superconducting phase upon pressure-induced interaction enhancement~\cite{ref3}. 

The resulting phase diagram shares qualitative similarities with other strongly correlated superconductors, including cuprates and organics, where a dome-shaped superconducting region appears adjacent to a Mott insulating phase. However, fullerides are distinguished by the fact that the superconducting state is predominantly s-wave in symmetry despite the strong electronic correlations. This unconventional coexistence of phonon-mediated pairing and Mott physics has motivated extensive theoretical efforts to uncover the microscopic mechanism underlying superconductivity and other correlated phases in these systems~\cite{ref4}.

Recent work on alkali-doped fullerides has continued to enrich our understanding of the interplay between strong correlations, lattice effects, and superconductivity in these molecular solids~\cite{ref5, ref5a, ref5b, ref5c, ref5d, ref5e, ref5f, ref5g, ref5h, ref5i, ref5j, ref5k, ref5l, ref5m, ref5n, ref5o, ref5p, ref5q, ref5r, ref5s, ref5t, ref5u, ref5v, ref5w, ref5Aa, ref5Ab, ref5Ac, ref5Ad, ref5Ae, ref5Af}. High-resolution angle-resolved photoemission spectroscopy on superconducting K$_3$C$_{60}$ thin films has revealed pronounced quasiparticle renormalization and strong electron-phonon coupling, evidenced by band kinks and replica features associated with high-energy Jahn–Teller active phonons, suggesting a strong-coupling nature of the superconductivity and indicating correlated electronic behavior beyond simple band-structure expectations~\cite{ref6}. Experimentally, extensive high-field studies have demonstrated that near the Mott insulating boundary the upper critical field of fulleride superconductors can reach values as high as 90 T, among the largest observed in cubic superconductors, with a concomitant crossover from weak- to strong-coupling superconductivity as the Mott transition is approached~\cite{ref3}. On the theoretical front, recent density functional theory plus dynamical mean-field theory investigations have revisited the stability and electronic structure of various polymorphs of  Cs$_3$C$_{60}$, clarifying that unconventional s-wave superconductivity emerges from a correlated metallic state adjacent to a Mott phase, with the nature of the superconductor–normal transition depending sensitively on lattice volume and pressure~\cite{ref7}. Additionally, theoretical studies have emphasized the importance of dynamical Jahn-Teller distortions and inverted Hund’s coupling in multiorbital models of fullerides, finding that such intramolecular interactions promote local pairing and give rise to exotic phases, including a Jahn–Teller metal with unusual normal-state properties and potentially enhanced superconducting coherence effects~\cite{ref8}. Together, these recent experimental and theoretical advances underscore the continued importance of correlated multiorbital and lattice physics in governing the rich phase behavior of alkali-doped fullerides, motivating further investigations across dimensionalities and lattice geometries. A growing consensus has emerged that the key to the fulleride phase diagram lies in the subtle interplay between intramolecular Jahn–Teller phonons and multiorbital Coulomb interactions, which together produce an effectively inverted Hund’s coupling. This unusual interaction hierarchy stabilizes low-spin molecular configurations and promotes local pairing tendencies, thereby enabling superconductivity near the Mott transition. Recent first-principles studies have successfully reproduced the experimentally observed phase diagram, including the close proximity of superconductivity and a low-spin Mott insulating phase, emphasizing the cooperative role of electron correlations and phonons in these materials~\cite{ref4}. Despite this progress, several aspects of the correlated phase behavior in fullerides remain incompletely understood. In particular, the role of lattice geometry and coordination number in shaping the competition between Fermi-liquid, magnetic, insulating, and superconducting phases has not been systematically explored. The BCC lattice, realized in Cs$_3$C$_{60}$, and related compounds, provides an ideal platform to address this question, as it combines reduced frustration with a distinct electronic density of states compared to its FCC counterpart. Motivated by these considerations, in this work we investigate an effective Hubbard model on the BCC lattice phenomenologically appropriate for alkali-doped fullerides, focusing on the emergence of superconducting and strongly correlated phases across a broad range of interaction strengths and temperatures.

The remainder of this paper is organized as follows. In Sec. II, we introduce the effective lattice model and discuss the choice of parameters relevant to alkali-doped fulleride compounds. Section III is devoted to the analysis of the normal–superconducting phase transition, where the model is approximated by an exactly solvable Hatsugai–Kohmoto model supplemented by a BCS-type pairing term. In Sec. IV, we focus on the strong-coupling regime and investigate correlation-driven phases by applying the spin-rotationally invariant slave-boson formalism after neglecting the weak pairing term. Finally, Sec. V summarizes our main results and presents our conclusions.

\section{II. Model Hamiltonian}

\begin{figure}
\includegraphics[width=\columnwidth]{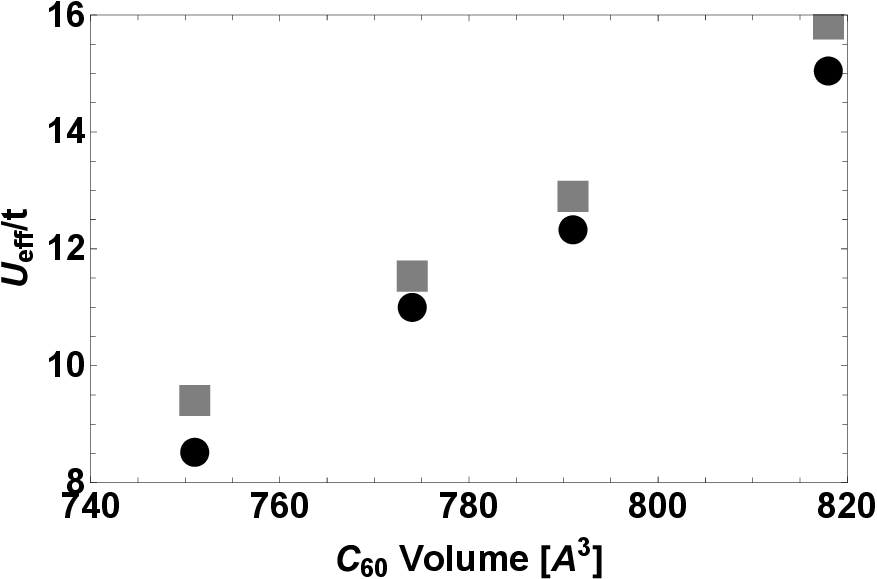}
\caption{Effective intra-orbital (black circles) and inter-orbital (gray squares) on-site interactions. These Coulomb interactions are renormalized due electron-phonon interactions.  For these calculation, the bare interaction parameters are taken from \emph{ab initio} calculations from ref.~\cite{ref9} and presented in ref.~\cite{ref5}.}\label{params}
\end{figure}

The lowest unoccupied molecular orbitals of the C$_{60}$ molecule is three-fold degenerate and the three electrons donated by the alkali-metal ions must occupy these orbitals. In general, the microscopic model must include density-density type interactions such as on-site and off-site Coulomb interactions, and non-density type interactions, such as the pair hopping and spin flip interactions. It has been shown that the off-site Coulomb interaction strength is about 25\% that of the on-site Coulomb interactions and the spin-flip interaction is estimated to be much smaller than that of the Coulomb interactions~\cite{ref9}. Further, it has been shown that the off-site Coulomb interaction and the spin-flip interaction do not play an essential role in driving the superconductivity~\cite{ref10}. In addition, the model must include the electron-phonon coupling term. In a recent publication, an effective microscopic model for the alkaline doped fullerides has been proposed~\cite{ref5}. The effective model consists with the kinetic energy, inter-orbital and intra-orbital Coulomb interactions, and pair-hopping interaction term \cite{ref5}. In the model derivation, the electron-phonon interactions has been eliminated using a standard perturbation theory in the anti adiabatic limit~\cite{ref1, ref12}. The model is supported by the recent bare interaction parameters calculated from \emph{ab} initio calculations~\cite{ref9} and the unconventional exotic behavior of the FCC structured Alkali-doped fullerides has been successfully explained \cite{ref5}. The explicit Hamiltonian for the model has the form,

\begin{eqnarray}
H = \sum_{\langle ij \rangle} \sum_m \sum_{\sigma}[t  - \mu \delta_{ij}] a^\dagger_{im\sigma} a_{jm\sigma}  \\ \nonumber
+ \frac{U_{eff}}{2} \sum_{i} [\sum_m a^\dagger_{im\sigma} a_{im\sigma} -3]^2 \\ \nonumber
+ J_{eff} \sum_{i} \sum_{m \neq m^\prime} a^\dagger_{im \uparrow} a_{im^\prime \uparrow} a^\dagger_{im \downarrow} a_{im^\prime \downarrow},
\end{eqnarray}

\noindent where $a^\dagger_{im\sigma}/a_{im\sigma}$ is the creation/annihilation operator for an electron in orbital $m$ at site $i$ with spin $\sigma$. The effective Coulomb interaction $U_{eff} = U + U_{ph}$ and the effective pair-hopping interaction $J_{eff} = J + J_{ph}$, where $U$ and $J$ are the bare on-site Coulomb interaction and bare Hund's coupling, have been renormalized due to the electron-phonon interactions. The parameters $U_{ph} < 0 $ and $J_{ph} < 0$ are the phonon contribution to the electron-electron interactions. For the FCC structured fulleride compounds, the effective interactions $U_{ph}$ and $J_{ph}$ are estimated for the entire experimentally relevant pressure range in Ref.~\cite{ref5}. This estimation is based on the recent \emph{ab initio} calculations for bare interaction parameters~\cite{ref9}. There in, it is concluded that both inter-orbital and intra-orbital effective Coulomb interaction $U_{eff} > 0$ is approximately equal and the effective Hund's coupling $J_{eff}$ is negative in the entire experimentally investigated pressure range. For the BCC structured alkali-doped fullerides, we find both inter-orbital and intra-orbital effective Coulomb interaction $U_{eff} > 0$  is  also approximately  equal  and  the  effective Hund’s coupling $J_{eff}$ is negative (see FIG.~\ref{params} and figure caption). In addition, there three major approximations are made when deriving this  model. First, the electron hopping between molecular orbitals at neighboring sites are assumed to be diagonal. Thus, the electron hopping between unlike molecular orbitals are neglected. Second, the off-site Coulomb interactions are neglected as those are shown to be only about $25\%$ that of on-site Coulomb interactions~\cite{ref9}. Third, the non-density type spin-flip interaction is also neglected as this does not play an essential role in driving the same orbital on-site s-wave superconductivity~\cite{ref10}. Usually the spin-flip term favors triplet superconductivity, however, the experiments support on-site spin-singlet superconductivity for the BCC structured alkali-doped fullerides.  

In the present paper, we plan to investigate the coexistence of superconductivity and antiferromagnetism of BCC structured lattice motivated by A15 fulleride compounds. The three competing terms in the Hamiltonian compete for three different phases. Let's denote the Hamiltonian $H = H_t + H_U + H_J$, where $H_t$ is the kinetic energy term that includes the chemical potential [first term in Eq. (1)], $H_U$ is the on-site repulsive term [second term in Eq. (1)], and $H_J$ is the on-site pair hopping term [third term in Eq. (1)]. While $H_t$ favors the metallic phase, $H_U$ and $H_J$ favor Mott-insulating and superconducting phases, respectively. For strong coupling limit where $U_{eff} \gg t$, the Mott-insulating phase tend to be antiferromagnetic. We consider the tunneling between only nearest neighbor pairs $\langle ij \rangle$ with tunneling amplitude $t$ and the chemical potential $\mu$. As the BCC lattice is a bipartite lattice, the BCC lattice can be divided in to two-sublattices $A$ and $B$, such that the atomic sites in each sub-lattices are only connected to the sites on the other sublattice through nearest neighbor connectors. For the BCC lattice, the atomic site at $d_l(0, 0, 0)$ has \emph{eight} nearest neighbors at $d_l (\pm 1/2, \pm 1/2 \pm 1/2)$ with $\sqrt{3} d_l/2$ distance away, where $d_l$ is the lattice constant.

\section{III. Superconductor- Normal Phase Transition}

Due to the nature of effective inverted Hunds coupling $J_{eff} < 0$ and the effective Coulomb interaction $U_{eff} > 0$ in our model, it is reasonable to expect a competition between superconductivity and other correlated phases in alkali-doped fulleride compounds. The on-site pair hopping term ($H_J$ ) favors the superconductivity, kinetic energy term ($H_t$) favors metallic behavior, and the on-site repulsive term ($H_U$) favors the correlated insulator or magnetic phase. 

\noindent First, Fourier transforming the fermionic operators using,

\begin{eqnarray}
a_{i m \sigma} = \frac{1}{\sqrt{N}} \sum_k e^{i \vec{k} \cdot \vec{r}_i} a_{k m \sigma},
\end{eqnarray}

\noindent for both sublattices and using the condition,

\begin{eqnarray}
\frac{1}{N} \sum_i e^{i (\vec{k} - \vec{k}^\prime) \cdot \vec{r}_i} = \delta_{k k^\prime},
\end{eqnarray}

\noindent we write the Hamiltonian terms,

\begin{eqnarray}
H_U = \frac{U_{eff}}{N} \sum_{m, m^\prime } \sum_{k, k^\prime} a^\dagger_{k m \uparrow} a_{k-q m \uparrow}  a^\dagger_{k^\prime m^\prime \downarrow} a_{k^\prime+q m^\prime \downarrow} 
\end{eqnarray}

\noindent and 

\begin{eqnarray}
H_J = \frac{J_{eff}}{N} \sum_{m, m^\prime} \sum_{k, k^\prime} a^\dagger_{k m \uparrow} a^\dagger_{-k m \downarrow}  a_{-k^\prime m^\prime \downarrow} a_{-k^\prime m^\prime \uparrow}.
\end{eqnarray}

\noindent Next, by defining the particle number operator $\hat{n}_{km} = a^\dagger_{k m \sigma} a_{k m \sigma}$, we approximate the Coulomb term by,

\begin{eqnarray}
H_U \simeq\frac{U_{eff}}{N} \sum_{m } \sum_{k} \hat{n}_{k m \uparrow} \hat{n}_{k m \downarrow}.
\end{eqnarray}

\noindent Note that this approximated term maximally breaks the hidden $Z_2$-symmetry ($\hat{n}_{km \uparrow} \rightarrow \hat{n}_{km \downarrow}$) of a Fermi liquid. Thus, this approximated Coulomb term represents a fixed point for Mott physics in renormalized sense~\cite{ref13}. This approximation is reasonable as our motivation is to study the superconducting transition in the presence of intermediate to strong effective on-site interaction range.

The Cooper instability towards s-wave superconducting state is included through the electron pairing in the $H_J$ term by defining the superconducting order parameter,

\begin{eqnarray}
\Delta = -\frac{J_{eff}}{N} \sum_k \langle a_{-k m \downarrow} a_{k m \uparrow} \rangle, 
\end{eqnarray}

\noindent where $\langle X \rangle$ is the thermal average of the operator $X$. Using a mean-field form of the $H_J$ term, our approximated Hamiltonian has the form $H = \sum_m H_{eff}$, where

\begin{eqnarray}
H_{eff} = \sum_{k \sigma} \xi_k  a^\dagger_{k \sigma} a_{k \sigma} + U_{eff} \sum_{k} \hat{n}_{k \uparrow} \hat{n}_{k \downarrow} \\ \nonumber - \sum_{k} \biggr(\Delta a^\dagger_{k \uparrow} a^\dagger_{-k \downarrow} + \Delta^\ast a_{-k \downarrow} a_{k \uparrow}\biggr) -\frac{N}{J_{eff}} |\Delta|^2. \label{HKBCS}
\end{eqnarray} 

\noindent Note that we can analyze $H_{eff}$ for the phase transition as the energy is independent of the degenerate orbital index $m$. As we are considering degenerate orbital, the sum over orbital index simply renormalizes the interaction parameters, thus we neglect the orbital index and treat $H_{eff}$ as $m$  independent. It has been argued that orbital degrees of freedom may play an important role in fullerides due to the relationship between the intraorbital interaction ($U$) and the interorbital interaction ($U^\prime$). However, within our approach, we find $U \approx U^\prime$, primarily due to phonon-mediated contributions. As a result, the orbital rotational invariance condition $U^\prime = U-2J$ is not applicable for the effective model in this context. For the same reason, we neglect the off-diagonal hopping contributions. The single particle band energy $\xi_k = t\sum_k e^{i \vec{k} \cdot \vec{\delta}} - \mu_{eff} = \epsilon_k - \mu_{eff}$, where $\vec{\delta}$ is the nearest-neighbor vector and $\mu_{eff}$ is the effective chemical potential. For the BCC lattice, the band energy dispersion $\epsilon_k = 8t [\cos(k_xd_l/2) \cos(k_yd_l/2) \cos(k_zd_l/2)]$ with lattice constant $d_l$. 

The final form of our Hamiltonian has the form $H_{eff} = H_{HK} + H_{SC}$, where $H_{HK}$ (first two terms in Eq.8) is the exactly-solvable Hatsugai-Kohomoto (HK) model~\cite{ref14} and $H_{SC}$ (second line in Eq. 8) is the BCS pairing interaction in the mean-field level. On the other hand, if one excludes the local interaction term in momentum space (second term in Eq. 8), our effective model is well-known Bardeen-Cooper-Schrieffer (BCS) model. The HK model is exactly solvable as the local interaction in the momentum space allows one to truncate the huge Hilbert space into a smaller one spanned by the basis states $\phi_k =|n_{k \uparrow}, n_{k \downarrow} \rangle$, where $n_{k \sigma} = 0, 1$ is the electron occupation number at k-state with spin $\sigma$. The recent studies suggest that the exactly solvable HK model provides an unconventional metallic ground state known as non-Fermi liquid state~\cite{ref15}. This is contrasts to the BCS model where the theory predicts a second-order phase transition from a conventional superconducting state to a Fermi-liquid metallic state. Therefore, it is reasonable to expect that our effective model referred as the HK-BCS mode $H_{eff}$ for BCC structured fulleride can give superconducting phase at the weak coupling limit, non-Fermi liquid metallic phase at intermediate coupling limit and a Mott-insulating phase at stronger coupling limit. Previous studies of the HK-BCS model on low dimensional lattices have demonstrated that, in the strong-coupling regime, the superconducting transition becomes first order rather than continuous~\cite{ref13, ref16, ref17, ref18}. In this section, we revisit the normal–superconducting phase transition within the HK-BCS framework, complementary to the strong-pairing analyses reported in Refs.~\cite{ref13} and~\cite{ref16}. In contrast to earlier studies largely focused on lower-dimensional systems, we construct the phase diagram in the context of a BCC lattice motivated by alkali-doped fullerides, thereby elucidating the role of three-dimensional lattice geometry and higher coordination on the nature of the transition.

The HK-BCS model can be exactly diagonalized in k-space span by the basis states $|\phi_k, \phi_{-k} >$ and Helmholtz free energy can be calculated exactly using the 16-eigenvalues of the Hamiltonian matrix~\cite{ref13, ref16}.

\begin{eqnarray}
F =-k_BT \ln(Z)
\end{eqnarray} 

\noindent where,

\begin{eqnarray}
Z = \sum_{n, k } e^{-\beta_T E_{k}^{n}}
\end{eqnarray} 

with $n =\{1, \cdot \cdot \cdot \cdot 16\}$ and $k$ runs only the half of the Brilouin zone. For $n =\{1, 2, 3\}$,

\begin{eqnarray}
E^{n}_{k} = 2 \xi_k +\frac{2U_{eff}}{3} + \frac{4}{\sqrt{3}} E^e_k \cos(\theta_k +\frac{2 \pi}{3} n),
\end{eqnarray} 

\noindent with 

\begin{eqnarray}
E^e_k = \sqrt{\biggr(\xi_k+\frac{U_{eff}}{2}\biggr)^2 + \Delta^2 +\frac{U_{eff}^2}{12}}
\end{eqnarray} 

\begin{eqnarray}
\theta_k = \frac{1}{3} \arccos \biggr[\frac{q_k}{(\sqrt{3} E^e_k)^3} \biggr]
\end{eqnarray} 

\begin{eqnarray}
q_k = U_{eff} \biggr(U_{eff}^2 + \frac{9}{2}(U_{eff} \xi_k + \xi_k^2)-\frac{9}{4} \Delta^2 \biggr)
\end{eqnarray} 

\begin{eqnarray}
E^4_k = 2 \xi_k +U_{eff}
\end{eqnarray} 
  
\begin{eqnarray}
E^5_k=2 \xi_k
\end{eqnarray} 

For $n = \{6,7\}$,

\begin{eqnarray}
E^n_k = 2 \xi_k + \frac{U_{eff}}{2} \mp E^o_k,
\end{eqnarray} 

\noindent with

\begin{eqnarray}
E^O_k = \sqrt{\biggr(\xi_k + \frac{U_{eff}}{2}\biggr)^2 + \Delta^2}
\end{eqnarray} 

\noindent with the lower sign is for  $n = 6$ and the upper sign is for $n = 7$. In evaluating the free energy, all 16 eigen states must be considered where the 16 states consist of non-degenerate $E^1_k$, $E^2_k$, and $E^3_k$, doubly degenerate $E^4_k$, triply degenerate $E^5_k$, and quarterly degenerate $E^6_k$ and $E^7_k$.

\begin{figure}
\includegraphics[width=\columnwidth]{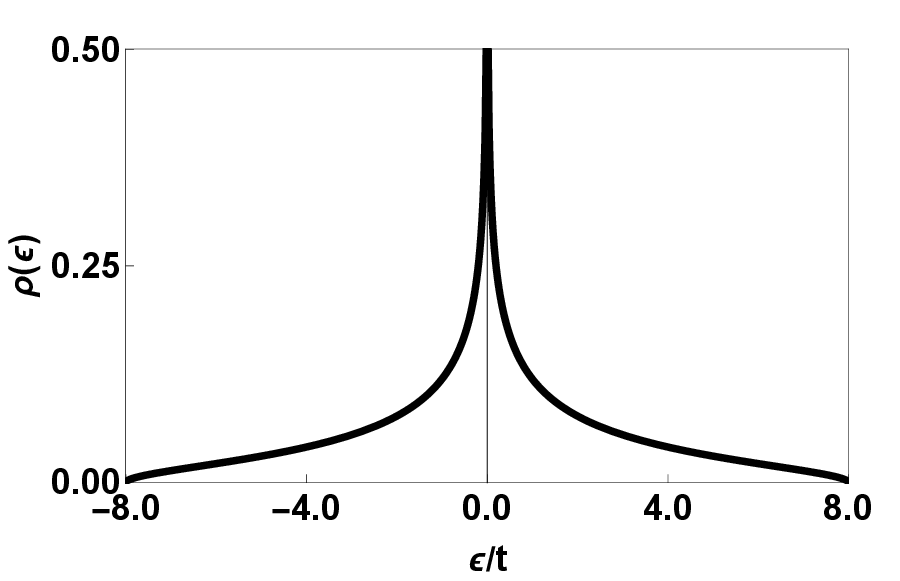}
\caption{Density of states for the nearest-neighbor only BCC lattice}\label{DOF}
\end{figure}

In evaluating the momentum sums appearing in our calculations, we convert the discrete sum over wave vectors in the first Brillouin zone into an energy integral using the density of states (DOS) of the underlying lattice. For a system defined on a body-centered cubic lattice with tight-binding dispersion $\epsilon_k$, the Brillouin-zone sum is replaced according to

\begin{eqnarray}
\frac{1}{N}\sum_{\mathbf{k}} (\cdots)
\;\longrightarrow\;
\int d\varepsilon\, \rho(\varepsilon)\, (\cdots),
\end{eqnarray}

\noindent where the single-particle density of states is defined as,

\begin{eqnarray}
\rho(\varepsilon)
=
\frac{1}{N}
\sum_{\mathbf{k}}
\delta(\varepsilon - \varepsilon_{\mathbf{k}}).
\end{eqnarray}

\noindent The DOS for the BCC lattice is computed numerically from the full band dispersion by sampling the first Brillouin zone and is normalized such that $\int d \epsilon \rho(\epsilon) = 1$. For the nearest-neighbor tight-binding model considered here, the energy spectrum is bounded within $\epsilon/t = \in [-8,8]$,  where $t$ denotes the hopping amplitude. The numerically obtained DOS is subsequently used in all our integrations. Numerically calculated, the DOS of the BCC lattice employed in this work is shown in Fig.~\ref{DOF}.

\begin{figure}
\includegraphics[width=\columnwidth]{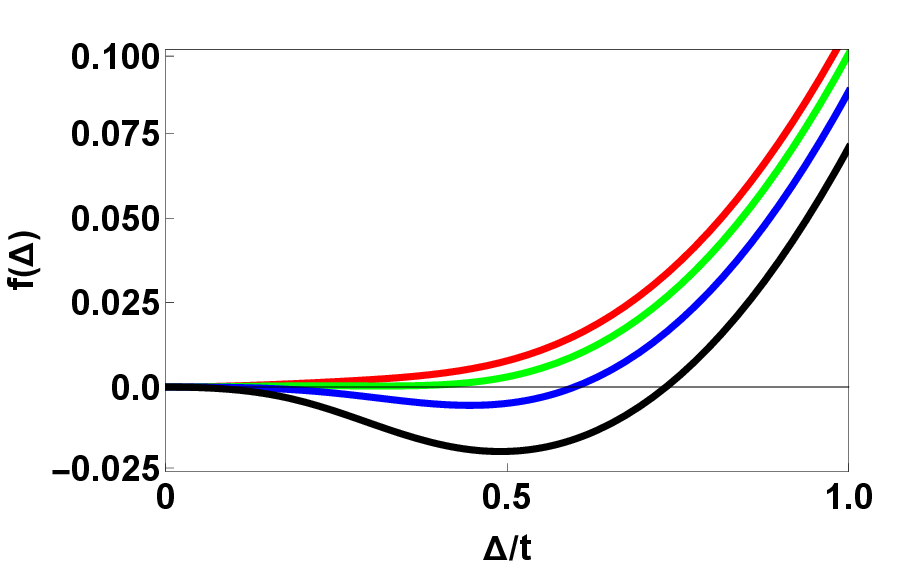}
\caption{Helmholtz free energy difference as a function of superconducting order parameter $\Delta$ at $U_{eff} = t$ and $J_{eff} = -2t$. As temperature (k$_BT$) decreases from $0.2t$ to $0.1t$ (Top to bottom: $0.2t$, $0.18t$, $0.15t$, $0.1t$), the secondary local minimum in the free energy becomes the global minimum indicating a first order phase transition from normal phase to superconducting phase.}\label{FE}
\end{figure}

\begin{figure}
\includegraphics[width=\columnwidth]{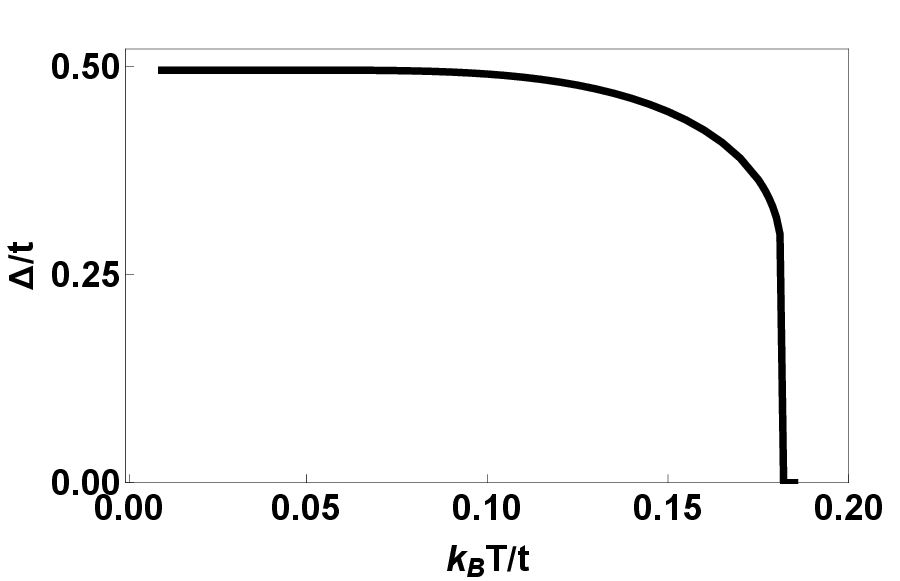}
\caption{Superconducting order parameter $\Delta/t$ as a function of temperature $k_BT/t$ at $U_{eff} = t$ and $J_{eff} = -2t$. The jump in the order parameter at $k_BT \simeq 0.18 t$ indicates the first order superconductor-Normal phase transition.}\label{PU1}
\end{figure}

\begin{figure}
\includegraphics[width=\columnwidth]{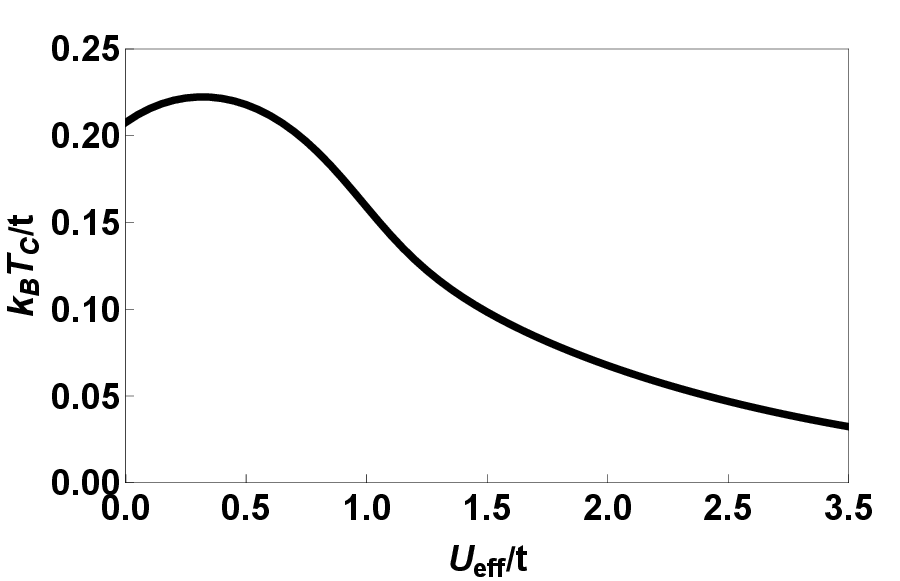}
\caption{Critical temperature for superconductor-Normal phase transition for $J_{eff} = -2t$. The normal-superconductor phase transition is of second order for $0 < U_{eff} < 0.3 t$ (upto maximum) second order and  it becomes first order for larger values of $U_{eff}$.}\label{SCCT}
\end{figure}

By evaluating the free energy as a function of the superconducting order parameter for a given set of model parameters, we first analyze the nature of the normal–superconducting phase transition. Figure~\ref{FE} illustrates the temperature evolution of the change in free energy per site $f(\Delta) = [F(\Delta)-F(0)]/(tN)$ as a function of the superconducting order parameter $\Delta$. At high temperatures, the free energy exhibits a single minimum at $\Delta = 0$, corresponding to a normal, non-Fermi-liquid state. Further lowering the temperature, a secondary local minimum develops at a finite value of $\Delta$, while the global minimum remains at $\Delta = 0$. As the temperature is reduced further, the finite-$\Delta$ minimum becomes the global minimum, signaling a discontinuous phase transition into the superconducting state. The coexistence of two local minima over an intermediate temperature range indicates that the normal–superconducting phase transition is first order for these parameters. This first order nature is further reflected in Fig.~\ref{PU1}, which shows the superconducting order parameter $\Delta$ as a function of temperature. We determined this $\Delta$ by numerically minimizing the free energy with respect to the superconducting order parameter. Consistent with the free-energy analysis, $\Delta$ exhibits a sharp jump at the critical temperature $T_C$, providing direct evidence for the first-order nature of the transition. The resulting phase boundary is summarized in Fig.~\ref{SCCT}, which displays the normal-superconductor critical temperature $T_C$ as a function of the effective on-site interaction $U_{eff}$ for a fixed inverted Hund’s coupling $J_{eff} = -2t$. For small values of $U_{eff}$, the critical temperature initially increases with increasing interaction strength, reaching a maximum near $U_{eff} \sim 0.5 t$. Beyond this point, $T_C$ decreases monotonically and eventually vanishes at a critical interaction strength $U_{eff} = U_C$, reflecting the suppression of superconductivity by strong local repulsion. The enhancement of $T_C$ at small $U_{eff}$ may be attributed either to the limited validity of the HK-BCS approximation in the weak-coupling regime or, alternatively, to a change in the nature of the phase transition for sufficiently small $U_{eff}$. Indeed we find that the normal–superconductor transition becomes second order  for $U_{eff} \sim 0.5 t$ at $J_{eff} = -2t$, consistent with a continuous evolution of the order parameter.

\section{IV.Fermi liquid, Mott insulating, and Antiferromagnetic phase transition}

To investigate the collective phases of alkali-fulleride in the strongly correlated regime, we employ the spin-rotationally invariant slave-boson (SRISB) approach to the effective Hubbard model on the BCC lattice. In the strong correlated regime away from the superconducting phase, the Hund’s coupling term is weaker, thus we neglected it in this part of the study. Further, we neglect the orbital degrees of freedom as the orbital effects simply renormalizes the interaction parameters due to the degenerate nature of the orbitals. Thus, in the strong correlation limit, our effective model on the BCC lattice in the real space has the form,   

\begin{eqnarray}
H = \sum_{\langle ij \rangle}  \sum_{\sigma}[t  - \mu \delta_{ij}] a^\dagger_{i\sigma} a_{j\sigma} + U_{eff} \sum_{i} a^\dagger_{i\uparrow} a_{i\uparrow} a^\dagger_{i\downarrow} a_{i\downarrow} 
\end{eqnarray}

The SRISB approach provides a controlled mean-field framework that captures the essential effects of local electronic correlations by explicitly separating charge and spin degrees of freedom, while preserving spin-rotational symmetry~\cite{ref19}. In this formulation, interaction effects lead to a renormalization of the quasiparticle weight and allow for the description of both itinerant and localized electronic behavior within a unified framework. As a result, the approach naturally interpolates between a weakly correlated Fermi liquid at small interaction strengths and a Mott insulating state at large interactions, and further permits the inclusion of spin symmetry breaking, such as antiferromagnetic order, at the saddle-point level. While fluctuations beyond mean field are neglected, the spin-rotational slave-boson theory has been shown to provide a reliable qualitative description of phase boundaries and correlation-driven transitions in Hubbard-type models for low dimensions, making it well suited for the present study~\cite{ref19}. In three-dimensional lattices, where spatial fluctuations are comparatively weaker, the saddle-point treatment underlying the slave-boson formalism is expected to be reasonably accurate. In the paramagnetic sector, the theory reproduces the Brinkman–Rice scenario for the Mott transition through the vanishing of the quasiparticle weight~\cite{ref20, ref21}.

The present formulation is closely connected to the original slave-boson approach introduced by Kotliar and Ruckenstein, which provides a transparent and physically intuitive description of correlation effects in the Hubbard model at the mean-field level~\cite{ref22}. While the Kotliar–Ruckenstein (KR) formalism successfully captures quasiparticle renormalization and the Brinkman–Rice picture of the Mott transition, its conventional implementation is not explicitly diagonal in a rotationally invariant Fock basis and is therefore limited in its ability to treat more general local interactions. At the saddle-point level, the KR approach is equivalent to the Gutzwiller approximation~\cite{ref23, ref24}, thereby establishing a direct connection between slave-boson mean-field theory and variational treatments of strong electronic correlations. The spin-rotationally invariant slave-boson approach employed here can be viewed as a natural extension of the KR scheme, restoring full spin-rotation symmetry and allowing for a consistent treatment of magnetic ordering and interaction terms beyond the density–density level, while retaining the essential physical insights of the original formulation.

In SRIRB theory, following ref.~\cite{ref19}, the original fermionic operators,

\begin{eqnarray}
a^\dagger_{i\sigma} = \sum_{\sigma^\prime} z^\dagger_{i, \sigma \sigma^\prime} f^\dagger_{i \sigma^\prime}
\end{eqnarray}

\noindent is written in terms of pseudofermionic operators $f^\dagger_{i \sigma}$ and bosonic operators $e_i$, $d_i$, and $p_{i \nu}$ with $\nu\in {0, 1, 2, 3}$. These bosonic operators label empty, doubly, and singly occupied states at site $i$. For each site,

\begin{eqnarray}
z_{i, \sigma \sigma^\prime} = e^\dagger_{i} p_{i, \sigma \sigma^\prime} +  \tilde{p}^\dagger_{i, \sigma \sigma^\prime}d_{i},
\end{eqnarray}

\noindent where  $p_{i, \sigma \sigma^\prime}$ are the matrix elements of the matrix, 

\begin{eqnarray}
P^\dagger_{i} = \frac{1}{2} \sum_{\nu = 0}^3 p^\dagger_{i, \mu} \underline{\tau}^\mu,
\end{eqnarray}

\noindent with $\underline{\tau}^{\mu = 0,1,2,3}$ are the Pauli matrices including the unit matrix $\underline{\tau}^{0}$. The time-reversal operator,

\begin{eqnarray}
\tilde{p}_{i, \sigma \sigma^\prime} = \frac{1}{2} \biggr(p_{i, 0} \tau^{0}_{\sigma \sigma^\prime} - \sum_{\nu = 0}^3 p^{\mu}_{i, \sigma \sigma^\prime}\biggr),
\end{eqnarray}

\noindent is necessary to account for the two spin states~\cite{ref19, ref20}. 

In mean field theory, the space-time dependent slave boson fields are replaced by their static expectation values, 

\begin{eqnarray}
e_i \rightarrow \langle e \rangle = e \\ \nonumber
d_i \rightarrow \langle d \rangle = d \\ \nonumber
p_{i 0} \rightarrow \langle p_0 \rangle = p_0 \\ \nonumber
p_{i \nu} \rightarrow \langle p_\nu \rangle = p 
\end{eqnarray}

with Lagrange’s multipliers,

\begin{eqnarray}
i \beta_{i 0} \rightarrow \langle \beta_0 \rangle = \beta_0 \\ \nonumber
i \alpha_{i} \rightarrow \langle \alpha \rangle = \alpha
\end{eqnarray}

\noindent to include the constraints,

\begin{eqnarray}
e^2 +d^2 + p_0^2 +p^2 =1 \\ \nonumber
2d^2 + p_0^2 + p^2 =1,
\end{eqnarray}

\noindent which guarantees that each site is occupied by one boson and average number of electrons per orbital is one at half filling. The effective chemical potential $\mu_{eff} = \mu - \beta_0$.

Following the standard path integral approach, the mean-filed free energy per lattice site at the saddle-point level is given by~\cite{ref25},

\begin{eqnarray}
f = -\frac{T}{N} \sum_{k, \pm} \ln \biggr[1+e^{-\beta_TE_{k,\pm}} \biggr] \\ \nonumber 
-\frac{U_{eff}}{2} \biggr( p_0^2 + p^2 -1\biggr) + \mu_{eff} - 2\beta p p_0
\end{eqnarray}

\noindent where,

\begin{eqnarray}
E_{k,\pm} = \frac{1}{4} \biggr\{\zeta \epsilon_{k, +} \pm \sqrt{\zeta^2 \epsilon^2_{k,-} + 16 \beta^2} \biggr\} - \mu_{eff}
\end{eqnarray}

\noindent and 

\begin{eqnarray}
\zeta = \frac{8p_0^2(1-p^2-p_0^2)}{1-4 p^2p_0^2} 
\end{eqnarray}

\noindent with $\epsilon_{k,\pm} = \epsilon_k \pm \epsilon_{k-Q}$. 

In the Mott insulating state $d = e = p = \beta = 0$ and $Q = (0, 0,0)$ lead to $\zeta =0$,  $p_0 = 1$, and  $E_{k, \pm} \rightarrow E^{Mott}_{k} = -\mu_{eff}$. These results the free energy  $f^{Mott} = -2 k_BT \ln(2)$.

In the Fermi-liquid phase, $p = \beta = 0$ and $Q = (0, 0,0)$ lead to $\zeta \rightarrow \zeta^{FL}  = 8 p_0^2 (1-p_0^2)$,  $p_0^2 = 1-2d^2$, and  $E_{k, \pm} \rightarrow E^{FL}_{k} = (\zeta^{FL} \epsilon_k)/2-\mu_{eff}$. Then the free energy in the Fermi liquid state becomes,

\begin{eqnarray}
F^{FL} = \frac{2k_BT}{N}\sum_k \ln \biggr[ 1+ e^{-\beta_T E_k^{FL}}\biggr] \\ \nonumber
- \frac{U_{eff}}{2}(p_0^2-1) +\mu_{eff}.
\end{eqnarray}

\noindent The saddle-point equations for $\mu_{eff}$ and $p_0$ for the Fermi liquid state are obtained by minimizing the free energy,

\begin{eqnarray}
\frac{2}{N}\sum_k f(\beta_T  E^{FL}_{k}) - 1 = 0,
\end{eqnarray}

\begin{eqnarray}
\frac{1}{N}\sum_k \epsilon_k f(\beta_T  E^{FL}_{k}) - \frac{U_{eff}}{16 (1-2p_0^2)} = 0,
\end{eqnarray}

\noindent where $f(x) = 1/(1 + e^{x})$ is the Fermi function.

In the AFM phase, $Q = (\pi, \pi,\pi)$ leads to $E_{k, \pm} \rightarrow E^{AFM}_{k, \pm} = \pm \frac{1}{2} \sqrt{\zeta^2_{AFM} \epsilon_k^2 + 4 \beta^2} - \mu_{eff}$ with,

\begin{eqnarray}
\zeta_{AFM} = \frac{8 p_0^2(1-p^2-p_0^2)}{1-4 p^2 p_0^2}.
\end{eqnarray}

\noindent The resulting saddle-point equations for the AFM state are given by,

\begin{eqnarray}
\frac{1}{N}\sum_{k,\pm} f(\beta_T  E^{AFM}_{k, \pm}) - 1 = 0,
\end{eqnarray}

\begin{eqnarray}
\frac{pp_0}{\beta} + \frac{1}{N}\sum_{k,\pm}  (\mp 1) \frac{f(\beta_T  E^{AFM}_{k, \pm})}{\sqrt{\zeta_{AFM}^2 \epsilon^2_K + 4 \beta^2 }} = 0,
\end{eqnarray}

\noindent and,

\begin{eqnarray}
\biggr(\frac{\zeta_{AFM}}{2} \frac{\partial \zeta_{AFM}}{\partial X}  \biggr)\frac{1}{N}\sum_{k,\pm} (\pm 1) \frac{ \epsilon^2_k f(\beta_T  E^{AFM}_{k, \pm})}{\sqrt{\zeta_{AFM}^2 \epsilon^2_k + 4 \beta^2 }}\\ \nonumber -( U_{eff}X + 2\beta \bar{X}) = 0,
\end{eqnarray}

\noindent where $X \in \{p, p_0\}$ and $\bar{X}$ denotes the complementary element,  $\bar{X} \in \{p_0, p\}$.

\begin{figure}
\includegraphics[width=\columnwidth]{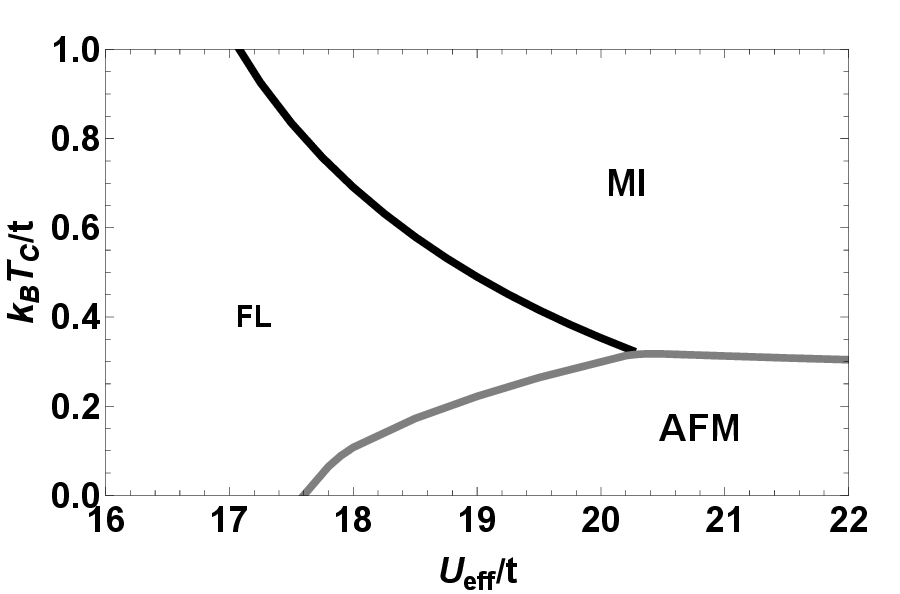}
\caption{Temperature -Interaction phase diagram for our effective Hubbard model on the BCC lattice. The spin-rotationally invariant slave-boson approach predicts three distinct phases, Fermi liquid (FL), Mott-insulating (MI), and antiferromagnetic (AFM) at the strong coupling limit at half filling.}\label{RISBPD}
\end{figure}

By converting the momentum sum into integral over the density of states as described before, we simultaneously solve our saddle-point equations numerically for given $U_{eff}$ for the phase boundary.  Figure~\ref{RISBPD} shows the temperature-interaction ($T-U$) phase diagram obtained within the spin-rotationally invariant slave-boson approach for our effective Hubbard model on the BCC lattice. Three distinct phases are identified: a paramagnetic Fermi-liquid (FL) phase at weak coupling, a Mott insulating (MI) phase at strong coupling, and an antiferromagnetic (AFM) phase at low temperatures and intermediate to strong interactions. Within the present mean-field treatment, all phase boundaries are found to be first order, consistent with the discontinuous evolution of the corresponding saddle-point solutions.

At low temperatures, the FL phase undergoes a transition into an AFM state upon increasing the on-site effective interaction strength. The FL–AFM transition at ground state occurs at a critical interaction $U_{c} \sim 17.6 t$ marking the onset of magnetic order driven by local moment formation. In a narrow intermediate temperature window near the multicritical region where FL, AFM, and MI phases meet, the system exhibits a sequence of transitions $FL\rightarrow AFM \rightarrow MI$ as one increase the interaction. This intermediate AFM region is confined to a very small temperature window, reflecting the strong competition between magnetic ordering and Mott localization close to multicritical point.

For larger values of $U$, the boundary between the AFM and MI phases varies weakly with increasing interaction strength, and the AFM ordering temperature is around $k_BT \sim 0.3 t$ beyond the multicritical point. In the strong-coupling limit, this scale can be interpreted in terms of an effective superexchange interaction arising from virtual hopping processes. For the Hubbard model, the antiferromagnetic exchange coupling is given by $J_{ex} = 4t^2/U_{eff}$, leading to a characteristic magnetic energy scale of order $zJ_{ex} S(S+1)/3$ on a lattice with coordination number $z$  and spin-$S$ (with $z = 8$ and $S = 1/2$ for the present study). This estimate gives the Neel's temperature $k_BT_N = zJ_{ex}/4 \sim 0.36 t$ for $U_{eff} =22 t$ which is consistent with our SRISB mean-field theory prediction of $k_BT_C \sim 0.3 t$. The magnitude of the ordering temperature extracted from the phase diagram is therefore consistent, up to mean-field perfectors, with the expected super exchange scale in the strong-coupling regime.

At higher temperatures, a first-order transition between the FL and MI phases is observed. The corresponding phase boundary does not terminate at large temperatures, reflecting the known limitation of the slave-boson saddle-point approximation in the high-temperature limit~\cite{ref20}. Nevertheless, low temperature regimes of primary interest here, the SRISB approach provides a coherent description of the interplay between itinerancy, Mott localization, and magnetic ordering on the BCC lattice, in qualitative agreement with established results for three-dimensional Hubbard systems~\cite{ref26, ref27, ref28, ref29}. The three phase structure for the BCC lattice shown in FIG.~\ref{RISBPD} closely resembles the phase diagram for the two dimensional square lattice constructed in Ref.~\cite{ref19}.

\section{V. Conclustions}

In this work, we studied the collective electronic phases of a fulleride-inspired effective Hubbard model on the body-centered-cubic lattice, motivated by alkali-doped fulleride molecular solids. The model incorporates renormalized on-site interactions and an effective inverted Hund’s coupling arising from electron–phonon interactions, allowing for a unified treatment of superconducting and correlation-driven phenomena within a lattice framework relevant to three-dimensional fullerides.

To access different interaction regimes, we employed two complementary theoretical approaches. In the intermediate-coupling regime, the on-site repulsive interaction was approximated by a long-range interaction in momentum space, leading to an exactly solvable Hatsugai–Kohmoto model supplemented by a BCS-type pairing term. Within this HK-BCS framework, we demonstrated a first-order normal–superconducting phase transition, characterized by a discontinuous jump of the superconducting order parameter and the coexistence of metastable states over a finite temperature range. The resulting nonmonotonic dependence of the superconducting critical temperature on the interaction strength highlights the nontrivial interplay between local repulsion and effective attractive interactions induced by the inverted Hund’s coupling.

In the strong-coupling regime, where pairing fluctuations are suppressed, we analyzed the same effective model using the spin-rotationally invariant slave-boson formalism. The resulting temperature–interaction phase diagram reveals first-order transitions between a paramagnetic Fermi-liquid phase, an antiferromagnetic phase, and a Mott insulating phase. In particular, we identified a narrow temperature window near the multicritical region where all three phases compete, reflecting the delicate balance between itinerancy, magnetic ordering, and Mott localization in three dimensions. At large interaction strengths, the antiferromagnetic ordering temperature saturates at a scale consistent with the expected superexchange interaction on the BCC lattice, supporting the physical interpretation of the magnetic phase within the strong-coupling limit.

Taken together, our results provide a coherent picture of how superconductivity, magnetism, and Mott physics emerge from an effective Hubbard-type description of alkali-doped fullerides. While both theoretical approaches rely on controlled approximations, namely the HK-BCS mapping in the intermediate-coupling regime and the saddle-point treatment inherent to the slave-boson formalism, the qualitative features of the phase diagrams and the competition between different ordered states are expected to be robust. Our study thus offers a unified framework for understanding the rich collective behavior of fulleride-inspired lattice systems and provides a useful starting point for future investigations incorporating fluctuation effects, multiband structure, or more quantitative treatments such as dynamical mean-field theory.

\end{document}